\documentclass[aps,prl,superscriptaddress,showpacs,floatfix,twocolumn]{revtex4}

\usepackage{amsmath}
\usepackage{graphicx}
\graphicspath{{./}}

\newcommand{\mean}[1]{\left\langle #1 \right\rangle} 
\newcommand{\jpsi}{J/\psi}
\newcommand{\pt}{{p_{\rm T}}}
\newcommand{\ncol}{N_{\rm coll}}
\newcommand{\npart}{N_{\rm part}}
\newcommand{\raa}{R_{\rm AA}}
\newcommand{\pp}{{p+p}}
\newcommand{\auau}{{\rm Au+Au}}

\newcommand{\dau}{d+{\rm Au}}

\begin{document}

\title{$\jpsi$ Production 
vs Centrality, Transverse Momentum, and Rapidity \\
in Au+Au collisions at $\sqrt{s_{NN}} = 200$~GeV
}

\newcommand{\abilene}{Abilene Christian University, Abilene, TX 79699, U.S.}
\newcommand{\banaras}{Department of Physics, Banaras Hindu University, Varanasi 221005, India}
\newcommand{\bnl}{Brookhaven National Laboratory, Upton, NY 11973-5000, U.S.}
\newcommand{\caucr}{University of California - Riverside, Riverside, CA 92521, U.S.}
\newcommand{\charlesczech}{Charles University, Ovocn\'{y} trh 5, Praha 1, 116 36, Prague, Czech Republic}
\newcommand{\ciae}{China Institute of Atomic Energy (CIAE), Beijing, People's Republic of China}
\newcommand{\cns}{Center for Nuclear Study, Graduate School of Science, University of Tokyo, 7-3-1 Hongo, Bunkyo, Tokyo 113-0033, Japan}
\newcommand{\colorado}{University of Colorado, Boulder, CO 80309, U.S.}
\newcommand{\columbia}{Columbia University, New York, NY 10027 and Nevis Laboratories, Irvington, NY 10533, U.S.}
\newcommand{\czechtech}{Czech Technical University, Zikova 4, 166 36 Prague 6, Czech Republic}
\newcommand{\dapnia}{Dapnia, CEA Saclay, F-91191, Gif-sur-Yvette, France}
\newcommand{\debrecen}{Debrecen University, H-4010 Debrecen, Egyetem t{\'e}r 1, Hungary}
\newcommand{\elte}{ELTE, E{\"o}tv{\"o}s Lor{\'a}nd University, H - 1117 Budapest, P{\'a}zm{\'a}ny P. s. 1/A, Hungary}
\newcommand{\fit}{Florida Institute of Technology, Melbourne, FL 32901, U.S.}
\newcommand{\fsu}{Florida State University, Tallahassee, FL 32306, U.S.}
\newcommand{\gsu}{Georgia State University, Atlanta, GA 30303, U.S.}
\newcommand{\hiroshima}{Hiroshima University, Kagamiyama, Higashi-Hiroshima 739-8526, Japan}
\newcommand{\ihepprot}{IHEP Protvino, State Research Center of Russian Federation, Institute for High Energy Physics, Protvino, 142281, Russia}
\newcommand{\illuiuc}{University of Illinois at Urbana-Champaign, Urbana, IL 61801, U.S.}
\newcommand{\instpasczech}{Institute of Physics, Academy of Sciences of the Czech Republic, Na Slovance 2, 182 21 Prague 8, Czech Republic}
\newcommand{\isu}{Iowa State University, Ames, IA 50011, U.S.}
\newcommand{\jinrdubna}{Joint Institute for Nuclear Research, 141980 Dubna, Moscow Region, Russia}
\newcommand{\kaeri}{KAERI, Cyclotron Application Laboratory, Seoul, South Korea}
\newcommand{\kek}{KEK, High Energy Accelerator Research Organization, Tsukuba, Ibaraki 305-0801, Japan}
\newcommand{\kfki}{KFKI Research Institute for Particle and Nuclear Physics of the Hungarian Academy of Sciences (MTA KFKI RMKI), H-1525 Budapest 114, POBox 49, Budapest, Hungary}
\newcommand{\korea}{Korea University, Seoul, 136-701, Korea}
\newcommand{\kurchatov}{Russian Research Center ``Kurchatov Institute", Moscow, Russia}
\newcommand{\kyoto}{Kyoto University, Kyoto 606-8502, Japan}
\newcommand{\labllr}{Laboratoire Leprince-Ringuet, Ecole Polytechnique, CNRS-IN2P3, Route de Saclay, F-91128, Palaiseau, France}
\newcommand{\lawllnl}{Lawrence Livermore National Laboratory, Livermore, CA 94550, U.S.}
\newcommand{\losalamos}{Los Alamos National Laboratory, Los Alamos, NM 87545, U.S.}
\newcommand{\lpc}{LPC, Universit{\'e} Blaise Pascal, CNRS-IN2P3, Clermont-Fd, 63177 Aubiere Cedex, France}
\newcommand{\lund}{Department of Physics, Lund University, Box 118, SE-221 00 Lund, Sweden}
\newcommand{\muenster}{Institut f\"ur Kernphysik, University of Muenster, D-48149 Muenster, Germany}
\newcommand{\myongji}{Myongji University, Yongin, Kyonggido 449-728, Korea}
\newcommand{\nagasaki}{Nagasaki Institute of Applied Science, Nagasaki-shi, Nagasaki 851-0193, Japan}
\newcommand{\newmex}{University of New Mexico, Albuquerque, NM 87131, U.S. }
\newcommand{\nmsu}{New Mexico State University, Las Cruces, NM 88003, U.S.}
\newcommand{\ornl}{Oak Ridge National Laboratory, Oak Ridge, TN 37831, U.S.}
\newcommand{\orsay}{IPN-Orsay, Universite Paris Sud, CNRS-IN2P3, BP1, F-91406, Orsay, France}
\newcommand{\peking}{Peking University, Beijing, People's Republic of China}
\newcommand{\pnpi}{PNPI, Petersburg Nuclear Physics Institute, Gatchina, Leningrad region, 188300, Russia}
\newcommand{\riken}{RIKEN, The Institute of Physical and Chemical Research, Wako, Saitama 351-0198, Japan}
\newcommand{\rikjrbrc}{RIKEN BNL Research Center, Brookhaven National Laboratory, Upton, NY 11973-5000, U.S.}
\newcommand{\rikkyo}{Physics Department, Rikkyo University, 3-34-1 Nishi-Ikebukuro, Toshima, Tokyo 171-8501, Japan}
\newcommand{\saispbstu}{Saint Petersburg State Polytechnic University, St. Petersburg, Russia}
\newcommand{\saopaulo}{Universidade de S{\~a}o Paulo, Instituto de F\'{\i}sica, Caixa Postal 66318, S{\~a}o Paulo CEP05315-970, Brazil}
\newcommand{\seoulnat}{System Electronics Laboratory, Seoul National University, Seoul, South Korea}
\newcommand{\stonybrkc}{Chemistry Department, Stony Brook University, Stony Brook, SUNY, NY 11794-3400, U.S.}
\newcommand{\stonycrkp}{Department of Physics and Astronomy, Stony Brook University, SUNY, Stony Brook, NY 11794, U.S.}
\newcommand{\subatech}{SUBATECH (Ecole des Mines de Nantes, CNRS-IN2P3, Universit{\'e} de Nantes) BP 20722 - 44307, Nantes, France}
\newcommand{\tenn}{University of Tennessee, Knoxville, TN 37996, U.S.}
\newcommand{\titech}{Department of Physics, Tokyo Institute of Technology, Oh-okayama, Meguro, Tokyo 152-8551, Japan}
\newcommand{\tsukuba}{Institute of Physics, University of Tsukuba, Tsukuba, Ibaraki 305, Japan}
\newcommand{\vandy}{Vanderbilt University, Nashville, TN 37235, U.S.}
\newcommand{\waseda}{Waseda University, Advanced Research Institute for Science and Engineering, 17 Kikui-cho, Shinjuku-ku, Tokyo 162-0044, Japan}
\newcommand{\weizmann}{Weizmann Institute, Rehovot 76100, Israel}
\newcommand{\yonsei}{Yonsei University, IPAP, Seoul 120-749, Korea}
\affiliation{\abilene}
\affiliation{\banaras}
\affiliation{\bnl}
\affiliation{\caucr}
\affiliation{\charlesczech}
\affiliation{\ciae}
\affiliation{\cns}
\affiliation{\colorado}
\affiliation{\columbia}
\affiliation{\czechtech}
\affiliation{\dapnia}
\affiliation{\debrecen}
\affiliation{\elte}
\affiliation{\fit}
\affiliation{\fsu}
\affiliation{\gsu}
\affiliation{\hiroshima}
\affiliation{\ihepprot}
\affiliation{\illuiuc}
\affiliation{\instpasczech}
\affiliation{\isu}
\affiliation{\jinrdubna}
\affiliation{\kaeri}
\affiliation{\kek}
\affiliation{\kfki}
\affiliation{\korea}
\affiliation{\kurchatov}
\affiliation{\kyoto}
\affiliation{\labllr}
\affiliation{\lawllnl}
\affiliation{\losalamos}
\affiliation{\lpc}
\affiliation{\lund}
\affiliation{\muenster}
\affiliation{\myongji}
\affiliation{\nagasaki}
\affiliation{\newmex}
\affiliation{\nmsu}
\affiliation{\ornl}
\affiliation{\orsay}
\affiliation{\peking}
\affiliation{\pnpi}
\affiliation{\riken}
\affiliation{\rikjrbrc}
\affiliation{\rikkyo}
\affiliation{\saispbstu}
\affiliation{\saopaulo}
\affiliation{\seoulnat}
\affiliation{\stonybrkc}
\affiliation{\stonycrkp}
\affiliation{\subatech}
\affiliation{\tenn}
\affiliation{\titech}
\affiliation{\tsukuba}
\affiliation{\vandy}
\affiliation{\waseda}
\affiliation{\weizmann}
\affiliation{\yonsei}
\author{A.~Adare}	\affiliation{\colorado}
\author{S.~Afanasiev}	\affiliation{\jinrdubna}
\author{C.~Aidala}	\affiliation{\columbia}
\author{N.N.~Ajitanand}	\affiliation{\stonybrkc}
\author{Y.~Akiba}	\affiliation{\riken} \affiliation{\rikjrbrc}
\author{H.~Al-Bataineh}	\affiliation{\nmsu}
\author{J.~Alexander}	\affiliation{\stonybrkc}
\author{A.~Al-Jamel}	\affiliation{\nmsu}
\author{K.~Aoki}	\affiliation{\kyoto} \affiliation{\riken}
\author{L.~Aphecetche}	\affiliation{\subatech}
% ...
\collaboration{PHENIX Collaboration} \noaffiliation


\begin{abstract}

The PHENIX experiment at the Relativistic Heavy Ion Collider (RHIC) has 
measured $\jpsi$ production for rapidities $-2.2 < y < 2.2$ in $\auau$ 
collisions at $\sqrt{s_{NN}}= 200$~GeV. The $\jpsi$ invariant yield and 
nuclear modification factor $\raa$ as a function of centrality, transverse 
momentum and rapidity are reported. A suppression of $\jpsi$ relative to 
binary collision scaling of proton-proton reaction yields is observed. 
Models which describe the lower energy $\jpsi$ data at the Super Proton 
Synchrotron (SPS) invoking only $\jpsi$ destruction based on the local 
medium density would predict a significantly larger suppression at RHIC 
and more suppression at mid rapidity than at forward rapidity. Both trends 
are contradicted by our data.

\end{abstract}



\maketitle




The Quark-Gluon-Plasma (QGP) is a state of deconfined quarks and gluons 
which is predicted by lattice Quantum Chromodynamics (QCD) calculations to 
be formed above a temperature of order $T_c = 170$~MeV for a baryon 
chemical potential $\mu_b=0$~\cite{Karsch:2001cy}. Heavy quarkonia 
($\jpsi$, $\psi'$, $\chi_c$ and $\Upsilon$) have long been considered one 
of the most promising probes to study formation and properties of QGP. In 
the deconfined state, the attraction between heavy quarks and anti-quarks 
is predicted to be reduced due to dynamic screening effects, leading to 
the suppression of heavy quarkonia yield. The strength of the suppression 
depends on the binding energies of the quarkonia and the temperature of 
the surrounding system~\cite{Matsui:1986dk}. Recent lattice QCD 
calculations suggest that the $\jpsi$ may not dissociate until well above 
$T_c$~\cite{Karsch:2002wv,Asakawa:2003re}. On the other hand $\chi_c$ and 
$\psi'$ which contribute to the total $\jpsi$ yield via decay are expected 
to dissolve at lower temperatures due to smaller binding energies.

A $\jpsi$ suppression observed at lower energies by the NA50 experiment at 
the SPS~\cite{Abreu:1997jh,Alessandro:2004ap} could be reproduced by 
various theoretical calculations, some invoking QGP formation and others 
not. A larger suppression is expected at RHIC compared to SPS due to the 
larger energy density of the medium 
created~\cite{Grandchamp:2003uw,Capella:2005cn}. On the other hand, 
several models predict that the $\jpsi$ yield will result from a balance 
between destruction due to thermal gluons and enhancement due to 
coalescence of uncorrelated $c\overline{c}$ 
pairs~\cite{Grandchamp:2003uw,Thews:2005vj}, which are produced abundantly 
in the initial collisions at RHIC 
energy~\cite{Adler:2005xv,phenix:2006hc}. Cold nuclear matter (CNM) 
effects such as nuclear absorption, shadowing and anti-shadowing are also 
expected to modify the $\jpsi$ yield. PHENIX $\dau$ data show that CNM 
effects are smaller at RHIC than those observed at lower 
energy~\cite{Adler:2005ph} and can be reproduced by a nuclear absorption 
cross-section of up to 3~mb plus nuclear shadowing~\cite{Vogt:2005ia}.

We report results on $\jpsi$ production measured by the PHENIX 
collaboration at mid-rapidity ($|y|<0.35$) via $e^+ e^-$ decay and at 
forward rapidity ($|y|\in[1.2,2.2]$) via $\mu^+\mu^-$ decay in $\auau$ 
collisions at $\sqrt{s_{NN}} = 200$~GeV. These results do not separate 
primordial $\jpsi$ and $\jpsi$ from $\chi_c$, $\psi'$ or $B$ decay. The 
$\jpsi$ invariant yields as a function of centrality, rapidity ($y$) and 
transverse momentum ($\pt$) are shown. They are combined with the $\jpsi$ 
yield measured in $\pp$ collisions~\cite{Adare:2007ph} to form the $\jpsi$ 
nuclear modification factor $\raa$.



The PHENIX apparatus is described in~\cite{Adcox:2003zm}. At mid-rapidity 
electrons are measured with two spectrometers consisting of Drift Chambers 
(DC), Pad Chambers (PC), Ring Imaging Cerenkov Counters (RICH), and 
Electromagnetic Calorimeters (EMCal). They are identified by matching 
charged tracks reconstructed with the DC and the first layer of the PC to 
clusters in the EMCal and to hits in the RICH. The energy-momentum matching 
requirement is $(E/p-1)\geq-2.5$ standard deviations ($\sigma$). The 
position matching between the track and the cluster in the EMCal is $\leq 
2.5\sigma$ ($4\sigma$) in azimuth and along the beam axis, for central 
(peripheral) collisions. For the RICH, at least 4 (2) matching hits are 
required. Muons are measured with two forward spectrometers consisting of 
a front absorber to stop most hadrons produced in the collision, cathode 
strip chambers (MuTr) which provide momentum information and a Muon 
Identifier (MuID) which uses alternating layers of steel absorber and 
Iarocci tubes. Charged particle trajectories are first reconstructed in 
the MuID then in the MuTr. They must reach the last plane of the MuID and 
have a good geometrical match between the MuID and the MuTr to be 
identified as muons. The matching is $<9^\circ$ for the slope and $<15$ 
(20)~cm for the position in the first layer of the MuID at positive 
(negative) rapidity. The collision centrality is determined using two 
Beam-Beam Counters (BBC) and Zero Degree Calorimeters 
(ZDC)~\cite{Adler:2003qi}.



The data used for this analysis were collected during the 2004 run at RHIC 
using a minimum bias trigger (a coincidence of the two BBC) which covers 
$92\pm3\%$ of the $\auau$ inelastic cross-section. After quality assurance 
and vertex cut $(|z|\leq30$~cm), $9.9\times10^8$ ($1.1\times10^9$) events 
were analyzed for mid (forward) rapidity, corresponding to an integrated 
luminosity of 157~$\mu$b$^{-1}$ (174~$\mu$b$^{-1}$). The forward rapidity 
data were filtered using an offline level-2 trigger which provides a fast 
reconstruction of the particle trajectory in the MuID. Events are accepted 
by this filter when at least two good quality tracks reaching the last 
plane of the MuID are found within the acceptance.



The $\jpsi$ yield is obtained from the unlike sign dilepton invariant 
mass distribution after subtracting the combinatorial background using an 
event-mixing technique. The background is normalized to the real data 
using the like-sign dilepton invariant mass distribution, 
$2\sqrt{N^{++}N^{--}}$, with $N^{++}$ ($N^{--})$ being the number of 
positive (negative) dilepton pairs. The accuracy of the normalization is 
estimated to be 2~\% and accounted for in the systematic errors. At 
midrapidity the $\jpsi$ mass resolution is $\sim35$~MeV/$c^2$. The number 
of $\jpsi$ is determined by counting the remaining unlike sign pairs in 
the mass range $2.9\leq M_{e^+e^-}\leq 3.3$~GeV/$c^2$. This number is 
corrected by the estimated contribution of the dielectron continuum and 
the loss due to the radiative tail. A total of $\sim 1000\;\jpsi$ are 
obtained and the signal to background (S/B) varies from 0.5 for central 
collisions to 15 for peripheral collisions. At forward rapidity, the 
$\jpsi$ mass resolution varies from 150 to 200~MeV/$c^2$ and is larger 
than at midrapidity primarily because of the multiple scattering and 
energy loss straggling in the front absorber. The residual background 
(notably from the open charm pairs and Drell-Yan) in the unlike-sign 
invariant mass distribution is evaluated using an exponential form. The 
$\jpsi$ signal is estimated with direct counting of the remaining pairs in 
the mass range $2.6\leq M_{\mu^+\mu^-}\leq 3.6$~GeV/$c^2$ and using a fit 
with different line shapes. The average of the resulting values is used as 
the number of $\jpsi$ and their dispersion is included in the systematic 
error. A total of $\sim 4500\;\jpsi$ are obtained and S/B varies from 0.2 
for central collisions to 3 for peripheral collisions.



The $\jpsi$ invariant yield in a given centrality, $\pt$ and $y$ bin is:
\begin{equation}
\label{eq:xsection}
\frac{B_{ll}}{2\pi \pt}\frac{d^2N_{\jpsi}}{d\pt dy} = 
\frac{1}{2\pi \pt}
\frac{N_{\jpsi}}{N_{evt}\Delta y\Delta \pt A\varepsilon}
\end{equation}
with $B_{ll}$ being the branching ratio for $\jpsi\rightarrow l^+l^-$; 
$N_{\jpsi}$ the number of $\jpsi$ measured in the centrality bin; 
$N_{evt}$ the corresponding number of events; $\Delta y$ the rapidity 
range; $\Delta \pt$ the transverse momentum range and $A\varepsilon$ the 
acceptance and efficiency correction for $\jpsi$. $A\epsilon$ is 
determined by full GEANT simulation. It decreases with the collision 
centrality due to overlapping hits in the RICH, EMCal and MuTr, leading to 
an increasing amount of mis-reconstructed tracks which are then rejected 
by the analysis cuts. This effect is evaluated by embedding simulated 
$\jpsi$ in real events. For the most central collisions the efficiency 
loss is 20~\% at mid rapidity and 75~\% (50~\%) at positive (negative) 
rapidity.

The nuclear modification factor in a given centrality, $\pt$ and $y$ bin is:
\begin{equation}
\raa = \frac{d^2N_{\jpsi}^{AA}/d\pt dy}
{\ncol d^2N_{\jpsi}^{pp}/d\pt dy}
\end{equation}
with $d^2N_{\jpsi}^{AA}/d\pt dy$ being the $\jpsi$ yield in $\auau$ 
collisions in the centrality bin, $\ncol$ the corresponding mean number of 
binary collisions and $d^2N_{\jpsi}^{pp}/d\pt dy$ the $\jpsi$ yield in 
$\pp$ inelastic collisions.


The systematic errors on the $\jpsi$ invariant yield 
(Table~\ref{tab:syst_error}) are grouped into three categories: point to 
point uncorrelated (type A) for which the points can move independently 
one from the other; point to point correlated (type B) for which the 
points can move coherently though not necessarily by the same amount; and 
global errors for which all points move by the same relative amount. 
Statistical and uncorrelated systematic errors are summed in quadrature 
and represented with vertical bars; correlated systematic errors are 
represented with boxes and different colors/symbols are used for forward 
and mid rapidity because they are independent; global systematic errors 
are quoted directly on the figures. For $\raa$, additional systematic 
errors are associated with uncertainties in the calculation of $\ncol$ (10 
to 28~\%) and the $\jpsi$ yield in $\pp$ (12~\% and 7~\% at mid and 
forward rapidity, respectively). On the other hand the systematic errors 
that are common to $\auau$ and $\pp$ cancel.

\begin{table}
\centering
\caption{\label{tab:syst_error}
Sources of systematic errors on the $\jpsi$ invariant yield. Columns 2 (3) 
are the average values at mid (forward) rapidity. When two values are 
given, the first (second) is for peripheral (central) collisions. Errors 
of type A (type B) are point to point uncorrelated (correlated).}
\begin{ruledtabular} \begin{tabular}{ccccc}
source&$|y|<0.35$&$|y|\in[1.2,2.2]$&type\\
\hline
signal extraction&6.5 to 9~\%&4 to 24~\%&A\\
acceptance&6~\%&10~\%&B\\
efficiency&4.5 to 8~\%&4 to 16~\%&B\\
run by run variation&4~\%&5~\%&B\\
input $y$, $\pt$ distributions&2~\%&4~\%&B\\
\end{tabular} \end{ruledtabular}
\end{table}



\begin{figure}[tb]
\includegraphics[width=1.0\linewidth]{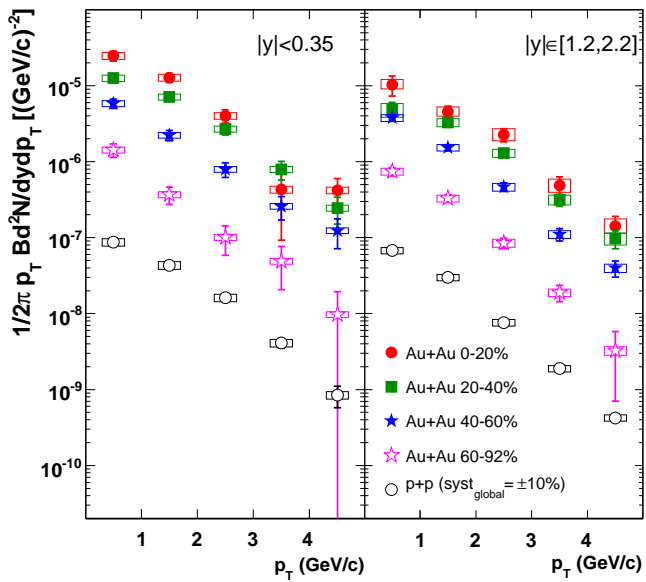}
\caption[]{\label{fig:yield_pt}
$\jpsi$ invariant yield vs. $\pt$ for different centrality bins in $\auau$ 
collisions and in $\pp$ collisions~\cite{Adare:2007ph}. The left (right) 
panel corresponds to mid (forward) rapidity. See text for description of 
the errors.}
\end{figure}

Figure~\ref{fig:yield_pt} shows the $\jpsi$ yield vs. $\pt$ for different 
centrality bins (see Table~\ref{table:meanpt} for the corresponding number 
of participants, $\npart$). Data from the two muon spectrometers are 
combined to obtain the forward rapidity points. In each centrality bin, 
the $\jpsi$ mean square transverse momentum, $\mean{\pt^2}$, is 
numerically calculated from the data for $\pt\le5$~GeV/$c$ and is shown in 
Table~\ref{table:meanpt}. The first error corresponds to the statistical 
and uncorrelated systematic error on the $\jpsi$ yield. The second 
corresponds to the correlated systematic error.  At midrapidity the 
$\mean{\pt^2}$ shows no variation versus centrality within the error bars. 
It increases slightly with $\npart$ at forward rapidity.
 
\begin{figure}[tb]
\includegraphics[width=1.0\linewidth]{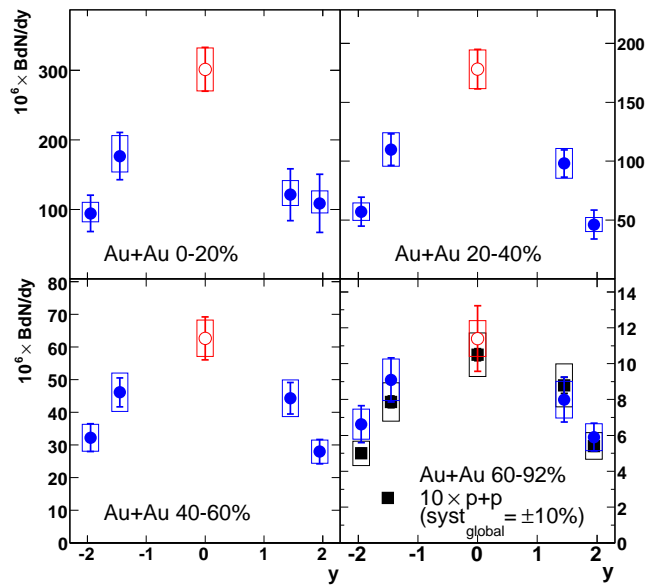}
\caption{\label{fig:yield_y} 
$\jpsi$ invariant yield vs. $y$ for different centrality bins in $\auau$ 
collisions and for $\pp$ collisions. Open (filled) circles are for mid 
(forward) rapidity $\auau$ data. Black squares are for $\pp$ 
data~\cite{Adare:2007ph}. See text for description of the errors. }
\end{figure}

Figure~\ref{fig:yield_y} shows the $\jpsi$ yield (integrated over $\pt$) 
vs. $y$ for different centrality bins. The root mean square (RMS) of each 
distribution is shown in Table~\ref{table:meanpt}. For the two most 
peripheral bins (40-60~\% and 60-93~\%) the RMS is compatible with that 
measured in $\pp$ collisions. For the most central bins (0-20~\% and 
20-40~\%), the RMS is smaller by about $2\sigma$.
 

\begin{table}
\centering
\caption{\label{table:meanpt} 
Characterization of the $\jpsi$ $\pt$ and $y$ distributions. Column 3 (4): 
$\jpsi$ $\mean{\pt^2}$ calculated for $\pt\le5$~GeV/$c$ at mid (forward) 
rapidity for different centrality bins in $\auau$ collisions and in $\pp$ 
collisions. The errors are descibed in the text.
Column 5: Calculated RMS of the 
corresponding $\jpsi$ $y$ distributions. }
\begin{ruledtabular}
\begin{tabular}{ccccc}
cent & $\npart$ &  $\langle \pt^2\rangle$ (GeV/$c$)$^2$ & $\langle \pt^2\rangle$ (GeV/$c$)$^2$ & $y$ RMS  \\
(\%) & &$|y|<0.35$	&$1.2<|y|<2.2$ & \\
\hline
0-20  & $280$ & $3.6 \pm  0.6 \pm  0.1$ & $  4.4 \pm   0.4 \pm   0.4$ & $ 1.32 \pm  0.06$\\
20-40 & $140$ & $4.6 \pm  0.5 \pm  0.1$ & $  4.6 \pm   0.3 \pm   0.4$ & $ 1.30 \pm  0.05$\\
40-60 & $ 60$ & $4.5 \pm  0.7 \pm  0.2$ & $  3.7 \pm   0.2 \pm   0.3$ & $ 1.40 \pm  0.04$\\
60-92 & $ 14$ & $3.6 \pm  0.9 \pm  0.2$ & $  3.3 \pm   0.3 \pm   0.2$ & $ 1.43 \pm  0.04$\\
$\pp$ & 2 & $4.1 \pm 0.2 \pm 0.1$ & $3.4\pm0.1\pm0.1$ & $ 1.41 \pm  0.03$ \\
\end{tabular}
\end{ruledtabular}
\end{table}

\begin{figure}[ht]
\includegraphics[width=1.0\linewidth]{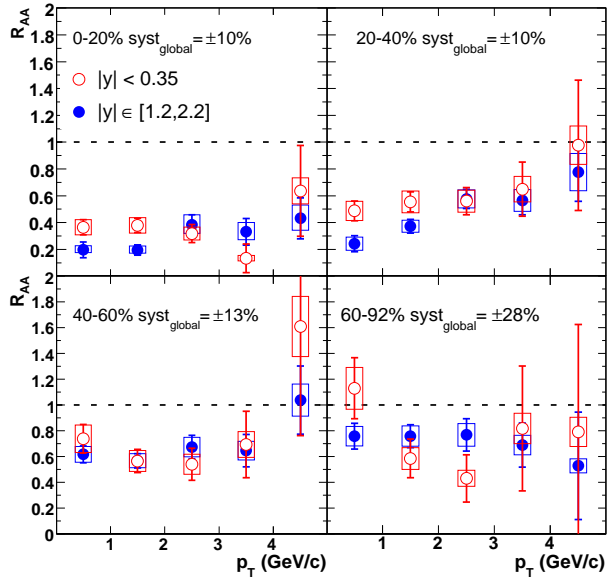}
\caption[]{\label{fig:r_aa_pt}
$\jpsi$ $\raa$ vs. $\pt$ for several centrality bins in $\auau$ collisions. Mid (forward) rapidity data are shown with open (filled) circles. See text for description of the errors.
}
\end{figure}

\begin{figure}[hb]
\includegraphics[width=1.0\linewidth]{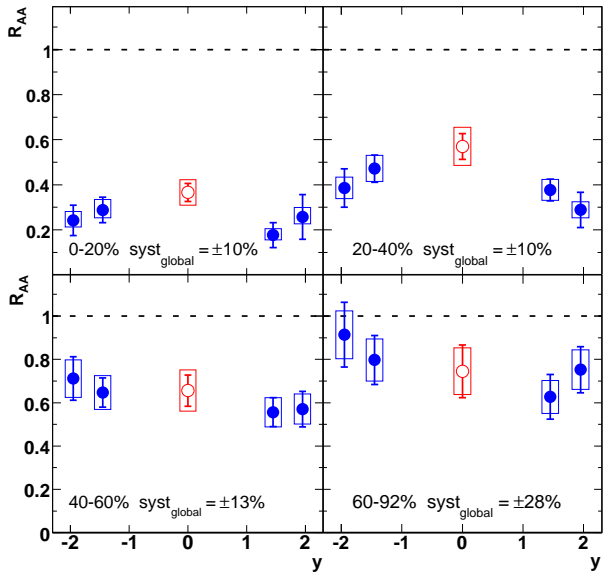}
\caption[]{\label{fig:r_aa_y}
$\jpsi$ $\raa$ vs. $y$ for different centrality bins. Open (filled) 
circles are for mid (forward) rapidity. See text for description of the 
errors.}
\end{figure}

\begin{figure}[bth]
\includegraphics[width=1.0\linewidth]{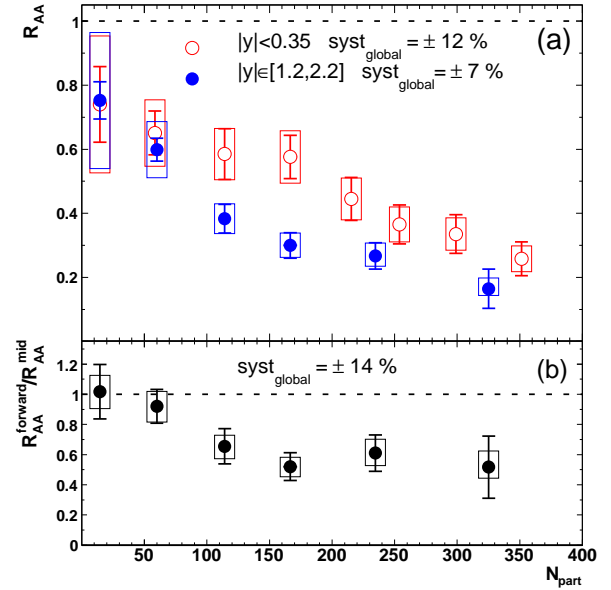}
\caption{\label{fig:r_aa_n_part} 
(a) $\jpsi$ $\raa$ vs. $\npart$ for $\auau$ collisions. Mid (forward) 
rapidity data are shown with open (filled) circles. (b) Ratio of  
forward/mid rapidity $\jpsi$ $\raa$ vs. $\npart$. For the two most 
central bins, mid rapidity points have been combined to form the ratio 
with the forward rapidity points. See text for description of the errors. 
}
\end{figure}

Figures~\ref{fig:r_aa_pt} and \ref{fig:r_aa_y} show the $\jpsi$ $\raa$ vs. 
$\pt$ and $y$, respectively, for different centrality bins. 
Figure~\ref{fig:r_aa_n_part}(a) shows the $\pt$ integrated $\raa$ vs. 
$\npart$ at mid and forward rapidity. For each rapidity, $\raa$ decreases 
with increasing $\npart$. For the most central collisions, $\raa$ is below 
0.3 (0.2) at mid (forward) rapidity. Figure~\ref{fig:r_aa_n_part}(b) shows 
the ratio of forward/mid rapidity $\raa$ vs. $\npart$. The 
ratio first decreases then reaches a plateau of about 0.6 for 
$\npart>100$.



We observed a significant $\jpsi$ suppression relative to binary scaling 
of proton-proton is observed for central $\auau$ collisions at RHIC. The 
magnitude of the suppression is similar to that observed at the 
SPS~\cite{Alessandro:2004ap} and greater than the suppression expected by 
extrapolating the cold nuclear matter effects measured in $\dau$ 
collisions~\cite{Vogt:2005ia,Adler:2005ph}. Models that describe the SPS 
data using a $\jpsi$ and/or $\chi_c$ and $\psi'$ suppression based on the 
local density predict a significantly larger suppression at RHIC than SPS 
and more suppression at mid rapidity than at forward 
rapidity~\cite{Grandchamp:2003uw,Capella:2005cn}. Both trends are 
contradicted by our data. Additionally, the $\jpsi$ mean square transverse 
momentum, restricted to $\pt\le5$~GeV/$c$, shows little dependence on 
centrality. Various models of $\jpsi$ production and suppression,
which predict very different transverse momentum and rapidity 
dependencies, can be significantly constrained by the data presented 
here and recent results on open charm~\cite{phenix:2006hc}.


We thank the staff of the Collider-Accelerator and 
Physics Departments at BNL for their vital contributions.  
We acknowledge support from 
the Department of Energy and NSF (U.S.A.), 
MEXT and JSPS (Japan), 
CNPq and FAPESP (Brazil), 
NSFC (China), 
MSMT (Czech Republic),
IN2P3/CNRS, and CEA (France), 
BMBF, DAAD, and AvH (Germany), 
OTKA (Hungary), 
DAE (India), 
ISF (Israel), 
KRF and KOSEF (Korea), 
MES, RAS, and FAAE (Russia),
VR and KAW (Sweden), 
U.S. CRDF for the FSU, 
US-Hungarian NSF-OTKA-MTA, 
and US-Israel BSF.

\end{document}